\documentclass{revtex4-2}
\usepackage[utf8]{inputenc}
\usepackage{amsbsy}
\usepackage{amsopn}
\usepackage{amstext}
\usepackage{amsmath,amsthm,amsfonts,amssymb}
\usepackage[mathcal]{eucal}
\usepackage{mathrsfs}
\usepackage{braket}
\usepackage[english]{babel}
\usepackage{color}
\usepackage{esint}
\usepackage[normalem]{ulem}
\usepackage{soul}
\usepackage{graphicx}
\usepackage{float}
\usepackage{units}
\usepackage{textcomp}
\usepackage{orcidlink}

\DeclareGraphicsExtensions{.png,PNG,.pdf,.PDF}
\usepackage{hyperref}
\usepackage{slashed}
\newcommand{\ie}{\begin{equation}}
\newcommand{\fe}{\end{equation}}
\newcommand{\se}{\begin{eqnarray}}
\newcommand{\ff}{\end{eqnarray}}
\begin{document}

\title{Dirac fermions with electric dipole moment and position-dependent mass in the presence of a magnetic field generated by magnetic monopoles}
\author{R. R. S. Oliveira\,\orcidlink{0000-0002-6346-0720}}
\email{rubensrso@fisica.ufc.br}
\affiliation{Departamento de F\'isica, Universidade Federal do Cear\'a (UFC), Campus do Pici, C.P. 6030, Fortaleza, CE, 60455-760, Brazil}


\date{\today}

\begin{abstract}

In this paper, we determine the bound-state solutions for Dirac fermions with electric dipole moment (EDM) and position-dependent mass (PDM) in the presence of a radial magnetic field generated by magnetic monopoles. To achieve this, we work with the $(2+1)$-dimensional (DE) Dirac equation with nonminimal coupling in polar coordinates. Posteriorly, we obtain a second-order differential equation via quadratic DE. Solving this differential equation through a change of variable and the asymptotic behavior, we obtain a generalized Laguerre equation. From this, we obtain the bound-state solutions of the system, given by the two-component Dirac spinor and by the relativistic energy spectrum. So, we note that such spinor is written in terms of the generalized Laguerre polynomials, and such spectrum (for a fermion and an antifermion) is quantized in terms of the radial and total magnetic quantum numbers $n$ and $m_j$, and explicitly depends on the EDM $d$, PDM parameter $\kappa$, magnetic charge density $\lambda_m$, and on the spinorial parameter $s$. In particular, the quantization is a direct result of the existence of $\kappa$ (i.e., $\kappa$ acts as a kind of ``external field or potential''). Besides, we also analyze the nonrelativistic limit of our results, that is, we also obtain the nonrelativistic bound-state solutions. In both cases (relativistic and nonrelativistic), we discuss in detail the characteristics of the spectrum as well as graphically analyze its behavior as a function of $\kappa$ and $\lambda_m$ for three different values of $n$ (ground state and the first two excited states).

\end{abstract}

\maketitle

\section{Introduction}

The so-called Dirac fermions are spin-1/2 massive particles governed by the Dirac equation (DE), which is a relativistic wave equation (equation of motion) developed by British physicist Paul Dirac in 1928, and is one of the fundamental equations of the Relativistic Quantum Mechanics (RQM) \cite{Dirac1,Dirac2,Greiner,Strange}. In that way, spin-1/2 massive particles (or only spin-1/2 particles) are also so-called Dirac particles (e.g., electrons, protons, neutrons, quarks, muons, taus, and possibly the neutrinos \cite{Studenikin,Lesgourgues}), that is, all (or almost all) fermions in the Standard Model of Particle Physics are Dirac fermions \cite{Martin,griffiths}. In particular, the DE satisfies the CPT and Lorentz symmetry/invariance, explains naturally the spin, helicity, and chirality of the spin-1/2 particles, and predicts that for each of these particles, there are their respective antiparticles (e.g., positrons, antiprotons, antineutrons, etc) \cite{Dirac1,Dirac2,Greiner,Strange,Studenikin,Lesgourgues,Martin,griffiths}. Besides, the DE (or Dirac fermions) has many applications (somewhat incalculable), for example, is used to model/study the good and ``old'' relativistic hydrogen atom \cite{Greiner,Strange}, Dirac oscillator \cite{Moshinsky,Franco,O0,O1,O2}, Aharonov-Bohm and Aharonov-Casher effects \cite{Aharonov,Hagen1}, quantum rings \cite{O4,O5}, ultracold and cold atoms \cite{Zhang2012}, trapped ions \cite{Lamata}, quantum Hall effect \cite{Schakel,O6}, quantum computing \cite{Huerta}, quantum phase transitions \cite{Bermudez1}, quark models \cite{Becirevic},  Hohenberg-Kohn-Sham density functional
formalism \cite{MacDonald}, quasiparticles (magnons, plasmons, and anyons) \cite{Manuel,Fransson,Pietro}, Dirac materials (e.g., graphene, fullerenes, metamaterials, semimetals, topological insulators, and carbon nanotubes) \cite{Novoselov,Gonzalez,McCann,Ahrens,Armitage,Huertas,Pietro}, relativistic quantum chemistry \cite{Visscher,Liu,Glushkov}, etc. Recently, the DE has been used to investigate the quantum Hall effect \cite{O7,O8}, Gödel-type spacetime \cite{O8,CQG}, rainbow gravity \cite{ARXIV2,ARXIV3}, electron-nucleus scattering \cite{Jakubassa}, non-Hermitian physics and strong correlation \cite{Yu}, BTZ black hole \cite{Guvendi}, etc.

In Particle Physics, a nonzero permanent electric dipole moment (NPEDM), or simply electric dipole moment (EDM), labeled by $\Vec{d}=\frac{2d}{\hbar}\Vec{s}=d\frac{\Vec{s}}{\vert\Vec{s}\vert}$, for a nondegenerate particle with spin $\Vec{s}=\frac{\hbar}{2}\Vec{\sigma}$ (i.e, Dirac fermions) directly implies the violation/breaking of time-reversal symmetry (T symmetry), i.e., an EDM is a direct signal of T-violation \cite{Abel,Chupp}. However, as the CPT symmetry (charge-parity-time symmetry) in any relativistic quantum field theories is invariant (CPT symmetry invariance), results that the observation of an EDM is also a signal of CP violation \cite{Abel,Chupp}. According to Ref. \cite{Chupp}, fermions with EDM have become a highly important focus of contemporary research for several interconnected reasons, such as: EDMs provide a direct experimental proof of CP violation (which is a feature of the Standard Model and Beyond-Standard-Model physics), the P-violating and T-violating EDM signal distinguishes the much weaker CP-violating interactions from the dominant strong and electromagnetic interactions, and the CP violation is a required component for the baryon asymmetry or matter-antimatter asymmetry (states that there is more matter than antimatter in the universe); therefore, new CP-violating interactions are required (i.e., EDMs could be one of the factors for explain this asymmetry). For more information about the EDM of atoms, molecules, nuclei (nucleons), and particles, we recommend Refs. \cite{Chupp,Commins,Pospelov,Engel}. Furthermore, EDMs have also been employed for a possible search for dark matter (i.e., a dark matter candidate could be Dirac fermions with EDM) \cite{Heo}. Recently, the DE (or Dirac fermions) with EDM has been used to study the noncommutative Dirac oscillator \cite{O9}, massive Dirac neutrinos \cite{EPjC}, axionic dark matter (or axion-fermion couplings) \cite{Smith,Luzio}, extended scotogenic models \cite{Fujiwara}, He–McKellar–Wilkens quantum phase \cite{Bakke}, electroweak baryogenesis \cite{Idegawa}, Higgs doublet model \cite{Hou}, and lepton flavor violation decays \cite{Tanimoto}.

In condensed matter physics (or solid-state physics), the concept of position-dependent mass (PDM), or spatially varying effective mass (SVEM), is very useful in studying the physical properties of various microstructures \cite{Alhaidari1}. For example, using the Schrödinger equation (SE) with PDM, we can investigate the electronic properties of semiconductors \cite{Bastard}, quantum wells and quantum dots \cite{Harrison,Serra}, 3He clusters \cite{Barranco}, quantum liquids \cite{Saavedra}, graded alloys and semiconductor heterostructures \cite{Roos1,Roos2,Morrow}, etc. In addition, exact solutions of the SE with PDM have already been obtained for several interesting potentials, such as the harmonic oscillator potential \cite{Jafarov}, hard-core potential \cite{Dong}, Hermitian and non-Hermitian potentials \cite{Jiang}, Hulthen potential \cite{Sever1}, modified Kratzer's molecular potential \cite{Sever2}, etc. However, the main difficulty in working with the SE with PDM consists of the problem of the ordering ambiguity in the kinetic term (or kinetic-energy operator), i.e., the ordering ambiguity of the mass and linear momentum operators (in this case, the mass no longer commutes with the momentum) \cite{Alhaidari1,Cavalcante}. Now, when we analyze the quantum dynamics of the particle with PDM through the DE, where the PDM $m(\Vec{r})$ and the momentum $\Vec{p}$ appear in the equation forming a linear combination whose coefficients are the Dirac matrices, this implies automatically in the elimination of the problem of the ordering ambiguity \cite{Alhaidari1,Cavalcante,Vakarchuk}. In this way, Dirac fermions with PDM also have a particular relevance from the point of view of RQM (or high-energy quantum systems). For example, using the DE with PDM, we can study problems involving the Coulomb potential \cite{Alhaidari1,Vakarchuk}, relativistic scattering \cite{Alhaidari2}, spin and pseudo-spin symmetry \cite{Ikhdair}, solid-state physics \cite{Renan}, infinite square well (or particle in a box) \cite{Alberto}, generalized gravitational uncertainty principle \cite{Pedram}, graphene \cite{Peres,Carmier,Jakubsk}, Aharonov-Bohm-Coulomb system \cite{O10,O11}, Aharonov-Casher effect \cite{O12}, etc. Recently, the DE with PDM was studied in curved spacetime backgrounds \cite{Ye}, and also with some interesting potentials, such as the Morse potential \cite{Bagchi}, modified manning Rosen potential \cite{Suparmi}, and the modified Woods Saxon potential \cite{Cari}.

This paper has as its goal to determine the bound-state solutions for (neutral) Dirac fermions with EDM and PDM in the presence of a magnetic field generated by magnetic monopoles. To achieve this, we work with the $(2+1)$-dimensional DE with nonminimal coupling in relativistic polar coordinates $(t,\rho,\theta)$, where the coupling constant is the EDM of the fermion. In particular, it is through the nonminimal coupling that the external magnetic field of the magnetic monopoles is introduced. However, to solve exactly and analytically the DE, we first convert it to a ``quadratic DE'', that is, to a second-order differential equation. In that way, we obtain the bound-state solutions (or the exact solutions for the bound-states of the fermionic system), given by the Dirac spinor (spinorial eigenfunction or spinorial wave function) and by the relativistic energy spectrum (or high-energy spectrum/levels), respectively. Besides, we also obtain the bound-state solutions for the nonrelativistic case (nonrelativistic bound-state solutions) or low-energy regime, given by the Schrödinger wave function and by the nonrelativistic energy spectrum (or low-energy spectrum/levels). So, with respect to the magnetic field of the magnetic monopoles, we consider the (inhomogeneous) radial magnetic field generated by a straight line (or linear distribution) of magnetic charges along the $z$-axis (i.e., perpendicular to the polar plane). Explicitly, this radial magnetic field (vector) is given as follows \cite{He,Wilkens,Chen,Dowling,B1,B2} {\color{red}(corrected)}
\begin{equation}\label{magneticfield}
\Vec{B}=\frac{\lambda_m}{\rho}\Vec{e}_\rho,
\end{equation}
where $\lambda_m=\lambda/2\pi$, being $\lambda\geq 0$ the (positive) magnetic charge per unit length, i.e., a magnetic charge linear density or simply magnetic charge density (for simplicity we will also call $\lambda_m$ of magnetic charge density), $\rho$ is the polar radial coordinate, given by $\rho=\sqrt{x^2+y^2}$ ($0\leq\rho<\infty$), and $\Vec{e}_\rho$ is the radial-unit vector, respectively. In particular, the magnetic field \eqref{magneticfield} is as if it were the magnetic analogue of the electric field of the Aharonov-Casher effect, where the field is generated by a straight line (linear distribution) of electric charges along the $z$-axis \cite{Aharonov,Hagen1,O0,O12}. It is interesting to comment that the magnetic monopole is (still) a hypothetical elementary particle that can be seen as an ``isolated magnet'' with only one magnetic pole (a north pole without a south pole or vice-versa), where the magnetic charge is one of its physical properties. Furthermore, the first quantum theory about magnetic monopoles emerged through a paper written by Dirac in 1931, i.e., a few years after he formulated his famous equation (therefore, the magnetic monopoles worked here are the Dirac monopoles) \cite{Dirac3,Dirac4}.

Now, with respect to the PDM of our problem, we consider the following variable mass \cite{O10,O11,O12,Alhaidari1,Vakarchuk,Alhaidari2,Ikhdair}
\begin{equation}\label{variablemass}
m(\rho)=m_0-\frac{\kappa}{\rho},
\end{equation}
where $m_0>0$ is the rest mass of the fermion, and $\kappa\geq 0$ is a positive real parameter (``mass parameter'' or ``PDM parameter''). In particular, the PDM \eqref{variablemass} can be seen as being a polar or two-dimensional version of a spherically symmetrical singular PDM worked in Refs. \cite{Alhaidari1,Vakarchuk,Alhaidari2,Ikhdair}. From a physical point of view, the parameter $\kappa$ can be defined (or interpreted) in two ways according to the literature. For example, $\kappa$ can be defined as $\kappa\equiv m_0\mu\lambdabar^2_C$, where $\mu$ is a real scale parameter with length inverse dimension, and $\lambdabar_C=\hbar/m_0 c$ is the (reduced) Compton wavelength of the fermion, where the rest mass (i.e., $m_0$) is obtained in the asymptotic limit ($\rho\to\infty$), or in the nonrelativistic limit ($\lambdabar_C\to 0$) of $m(\rho)$ \cite{Alhaidari1,Alhaidari2}. Then, a possible application for this singular mass term (i.e., $\kappa/\rho$) could be found in relativistic quantum field theory (QFT), which could contribute to the ``renormalization'' of ultraviolet divergences that occur at high energies (equivalently, small distances $\rho\to 0$ where $m(\rho)\to-\infty$, i.e., a negative divergent mass). Therefore, in such a model, these divergences are renormalized into the singular mass term where $\mu$ plays the role of the renormalization scale. In addition, $\kappa$ can also be defined as $\kappa\equiv m_0 a$, where the parameter $a$ with length dimension (formed by the mechanism of the particle interaction with the vacuum fluctuations in the presence of the force center) is related to the classical electron radius (Lorentz radius or Thomson scattering length) in the form $0\leq a<r_c$ ($r_c=e^2/4\pi\epsilon_0 m_e c^2$) \cite{Vakarchuk}. In this case, both the mass and the electrical charge of the fermion would be functions of the position. One justification for this is that interaction processes at QFT (subatomic scales) deform the Coulomb interaction in the atom when the distance between the electron and the nucleus is very small. Furthermore, it is interesting to comment that unlike \cite{O10,O11,O12,Alhaidari1,Vakarchuk,Alhaidari2,Ikhdair} (in which it works with the Coulomb potential and the variable mass term is positive), here, we consider the variable mass term to be negative (i.e., $-\kappa/\rho$). In fact, we did this to have a ``Coulomb-like attractive potential'' in our problem (such as occurs in \cite{O10,O11,O12,Vakarchuk,Ikhdair}) and, consequently, to obtain the bound states (spectrum) of the system.

This paper is organized as follows. In Sect. \ref{sec2}, we initially introduce the linear DE, or simply the DE, with nonminimal coupling for a fermion with EDM and PDM. Defining a new spinor for the system, we convert this DE to a quadratic DE. In order to simplify the calculations, that is, eliminate $\cos{\theta}$ and $\sin{\theta}$ of the equation, we use for such a similarity transformation given by a unitary operator $U(\theta)$. In Sect. \ref{sec3}, we solve exactly and analytically the second-order differential equation of our problem via a change of variable and the asymptotic behavior, where we obtain a generalized Laguerre equation. From this equation, we obtain the relativistic bound-state solutions of the system (Dirac spinor and relativistic energy spectrum). In particular, we discuss in detail the characteristics of this spectrum as well as graphically analyze the behavior of the spectrum as a function of the PDM parameter $\kappa$ and of the magnetic charge density $\lambda_m$ for three different values of $n$. In Sect. \ref{sec4}, we obtain the nonrelativistic bound-state solutions, given by the Schrödinger wave function and by the nonrelativistic energy spectrum. Finally, in Sect. \ref{sec5} we present our conclusions.

\section{The Dirac equation with electric dipole moment and position-dependent mass in the $(2+1)$-dimensional Minkowski spacetime \label{sec2}}

The $(3+1)$-dimensional covariant DE with nonminimal coupling that governs the relativistic quantum dynamics of a massive spin-1/2 fermion with EDM and PDM is given as follows (in SI units) \cite{O9,Commins,Greiner,O12}
\begin{equation}\label{dirac1}
\left(\gamma^\mu p_\mu-\frac{id}{2c}\sigma^{\mu\nu}\gamma_5F_{\mu\nu}-m(\vec{r})c\right)\Psi_D(t,\vec{r})=0, \ \ (\mu,\nu=0,1,2,3),
\end{equation}
where $\gamma^{\mu}=(\gamma^0,\vec{\gamma})$ are the gamma matrices, which satisfies the anticommutation relation of the Clifford Algebra: $\{\gamma^{\mu},\gamma^{\nu}\}= 2g^{\mu\nu}$, being $g^{\mu\nu}$=diag$(+1,-1,-1,-1)$ the Minkowski metric tensor (or Minkowski metric), $p_\mu=i\hbar\partial_\mu=(i\hbar\partial_0,i\hbar\partial_i)=(\frac{i\hbar}{c}\partial_t,i\hbar\partial_i)=(p_0,-\Vec{p})$ is the relativistic momentum operator (or momentum four-vector), being $\Vec{p}=-i\hbar\Vec{\nabla}$ the usual or canonical momentum operator (or nonrelativistic momentum operator), $d$ ($d=\pm \vert d\vert$) is the EDM of the fermion with a PDM given by $m(\Vec{r})$, $\sigma^{\mu\nu}=\frac{i}{2}[\gamma^\mu,\gamma^\nu]$ is an antisymmetric tensor (``Dirac tensor''), $\gamma_5$ is the fifth gamma matrix, $F_{\mu\nu}=\partial_\mu A_\nu-\partial_\nu A_\mu$ is the electromagnetic field tensor (or field strength), being $A_\mu=(A_0/c,-\Vec{A})$ the external electromagnetic field (or potential four-vector), $\Psi_D(t,\Vec{r})$ is the four-component Dirac spinor (or four-element column vector), and the constants $c$ and $\hbar$ are the speed of light and reduced Planck constant ($\hbar=h/2\pi$), respectively.

Now, let us obtain the DE in quadratic form, that is, a quadratic DE from the linear DE given in \eqref{dirac1}. In particular, this is an effective strategy for working with fermions with PDM \cite{O10,O11,O12,Vakarchuk} or constant mass {\color{red}\cite{O9,O1,Gavrilov,Chen2}} in external electromagnetic fields. Therefore, defining the spinor $\Psi_D$ in terms of a new spinor $\Phi$ as follows
\begin{equation}\label{dirac2}
\Psi_D(t,\vec{r})\equiv O_D\Phi(t,\vec{r})=\left(\gamma^\alpha p_\alpha-\frac{id}{2c}\sigma^{\alpha\beta}\gamma_5F_{\alpha\beta}+m(\vec{r})c\right)\Phi(t,\vec{r}), \ \ (\alpha,\beta=0,1,2,3),
\end{equation}
we then obtain the following quadratic DE
\begin{equation}\label{dirac3}
\left(\gamma^\mu p_\mu-\frac{id}{2c}\sigma^{\mu\nu}\gamma_5F_{\mu\nu}-m(\vec{r})c\right)\left(\gamma^\alpha p_\alpha-\frac{id}{2c}\sigma^{\alpha\beta}\gamma_5F_{\alpha\beta}+m(\vec{r})c\right)\Phi(t,\vec{r})=0,
\end{equation}
where $O_D$ can be seen as a ``Dirac operator'' (and has a form very similar to linear DE, however, with the mass term having the sign changed).

So, developing the indexes $\mu$ and $\nu$ of Eq. \eqref{dirac3}, we have \cite{O9}
\begin{equation}\label{dirac4}
A^-A^+\Phi(t,\vec{r})=0,
\end{equation}
where we define the operators $A^{\mp}$ as follows
\begin{equation}\label{operators}
A^\mp\equiv\left[\gamma^0\left(\frac{i\hbar}{c}\frac{\partial}{\partial t}\right)-\Vec{\gamma}\cdot(\Vec{p}+id\gamma^0\Vec{B})+\frac{d}{c}\vec{\Sigma}\cdot\Vec{E}\mp m(\vec{r})c\right],
\end{equation}
being $\Vec{B}=\Vec{B}_{ext}$ and $\Vec{E}=\Vec{E}_{ext}$ the external magnetic and electric fields.

However, in polar coordinates $(t,\rho,\theta)$ where $p_z=z=0$ \cite{O9,Andrade} (i.e., in $(2+1)$-dimensions), and considering only the magnetic field (i.e., $\Vec{B}\neq 0$ and $\Vec{E}=\Vec{0}$), Eq. \eqref{dirac4} becomes
\begin{equation}\label{dirac5}
A^-A^+\Phi(t,\rho,\theta)=0,
\end{equation}
where
\begin{equation}\label{operators2}
A^\mp\equiv\left[\gamma^0\left(\frac{i\hbar}{c}\frac{\partial}{\partial t}\right)+i\gamma^\rho\left(\hbar\frac{\partial}{\partial\rho}+d\gamma^0 B(\rho)\right)+\gamma^\theta\left(\frac{i\hbar}{\rho}\frac{\partial}{\partial\theta}\right)\mp m(\rho)c\right],
\end{equation}
being $\gamma^\rho=\Vec{\gamma}\cdot\Vec{e}_\rho=\gamma^1\cos{\theta}+\gamma^2\sin{\theta}$ and $\gamma^\theta=\Vec{\gamma}\cdot\Vec{e}_\theta=-\gamma^1\sin{\theta}+\gamma^2\cos{\theta}$ the projections or components of $\Vec{\gamma}$ (``gamma vector'') on the polar plane, $0\leq\theta\leq 2\pi$ is the angular coordinate (or polar angle), and we use $\vec{B}=B(\rho)\Vec{e}_\rho$.

On the other hand, due to the presence of $\cos{\theta}$ and $\sin{\theta}$, it is difficult to proceed without a simplification of Eq. \eqref{dirac5}. In that way, to eliminate this obstacle (i.e., $\cos{\theta}$ and $\sin{\theta}$ in equation), we can apply a similarity transformation whose unitary matrix/operator is given by $U(\theta)=e^{-i\theta\Sigma^3/2}\in SU(2)$, where $\Sigma^3=i\gamma^1\gamma^2$, $\Sigma^3\gamma^0=\gamma^0\Sigma^3$ and $U^{\dagger}U=UU^\dagger=1$ ($U^\dagger=U^{-1}$ or $(U^\dagger)^{-1}=U$) \cite{O1,O10,O11,O12}. In particular, this unitary matrix has the function of reducing the ``rotated or transformed matrices'' $\gamma^\rho$ and $\gamma^\theta$ to the fixed or Cartesian matrices $\gamma^1$ and $\gamma^2$ (which we know how to work easily) in the following form
\begin{equation}\label{spinor}
U^\dagger(\theta)\gamma^\rho U(\theta)=\gamma^1, \ \ U^\dagger(\theta)\gamma^\theta U(\theta)=\gamma^2.
\end{equation}

Therefore, applying a similarity transformation to Eq. \eqref{dirac5}, and using the magnetic field \eqref{magneticfield} and the PDM \eqref{variablemass}, we obtain
\begin{equation}\label{dirac6}
B^-B^+\phi(t,\rho,\theta)=0,
\end{equation}
where
\begin{equation}\label{operators3}
B^\mp\equiv\left[\gamma^0\left(\frac{i\hbar}{c}\frac{\partial}{\partial t}\right)+i\gamma^1\left(\hbar\frac{\partial}{\partial\rho}+\frac{\hbar}{2\rho}+\frac{d\lambda_m}{\rho} \gamma^0\right)+\gamma^2\left(\frac{i\hbar}{\rho}\frac{\partial}{\partial\theta}\right)\mp\left(m_0 c-\frac{\kappa c}{\rho}\right)\right],
\end{equation}
or better
\begin{equation}\label{operators4}
B^\mp\equiv\left[\sigma_3\left(\frac{i\hbar}{c}\frac{\partial}{\partial t}\right)+\sigma_2\left(\hbar\frac{\partial}{\partial\rho}+\frac{\hbar}{2\rho}\right)+i\sigma_1\left(\frac{i\hbar}{\rho}\frac{\partial}{\partial\theta}+\frac{d\lambda_m}{\rho}\right)\mp\left(m_0 c-\frac{\kappa c}{\rho}\right)\right],
\end{equation}
being $\phi(t,\rho,\theta)\equiv U^\dagger\Phi(t,\rho,\theta)$ the ``rotated or transformed spinor'', and we use the fact that we are working in the $(2+1)$-dimensional Minkowski spacetime (or planar spacetime), where we must write/define the gamma matrices $\gamma^\mu=(\gamma^0,\gamma^1,\gamma^2)=(\gamma_0,-\gamma_1,-\gamma_2)$ in terms of the $2\times 2$ Pauli matrices, i.e., $\gamma_0=\sigma_3$, $\gamma_1=\sigma_3\sigma_1=i\sigma_2$, and $\gamma_2=\sigma_3\sigma_2=-i\sigma_1$ \cite{O1,O10,O11,O12,Andrade}. Besides, the spinors $\phi(t,\rho,\theta)$ and $\Phi(t,\rho,\theta)$ must satisfy the following periodicity conditions: $\phi(t,\rho,\theta\pm 2\pi)=-\phi(t,\rho,\theta)$, and $\Phi(t,\rho,\theta\pm 2\pi)=\Phi(t,\rho,\theta)$.

Explicitly, we can obtain from Eq. \eqref{dirac6} a second-order differential equation, given by
\begin{equation}\label{dirac7}
\left[\frac{\partial^2}{\partial\rho^2}+\frac{1}{\rho}\frac{\partial}{\partial\rho}-\frac{1}{\rho^2}\left(\Gamma^2-\Gamma+\frac{1}{4}\right)+\frac{1}{\rho}\left(\frac{2m_0 c^2\kappa}{\hbar^2}\right)-\Delta\right]\phi(t,\rho,\theta)=0
\end{equation}
where we define the following operators
\begin{equation}
\Gamma^2\equiv\left(\frac{J_z}{\hbar}-\frac{d\lambda_m}{\hbar}\right)^2+\left(\frac{\kappa c}{\hbar}\right)^2, \ \ \Gamma\equiv\sigma_3\left(\frac{J_z}{\hbar}-\frac{d\lambda_m}{\hbar}\right)+\sigma_2\left(\frac{\kappa c}{\hbar}\right), \ \ \Delta\equiv\left(\frac{1}{c^2}\frac{\partial^2}{\partial t^2}+\frac{m^2_0 c^2}{\hbar^2}\right),
\end{equation}
being $J_z=-i\hbar\frac{\partial}{\partial\theta}$ the projection or component of the total angular momentum operator $\Vec{J}$ on the $z$-axis.

So, assuming that our system is a stationary quantum system (does not explicitly depend on time), implies that we can define an ansatz for the spinor $\phi(t,\rho,\theta)$ as follows \cite{O1,O10,O11,O12}
\begin{equation}\label{spinor}
\phi(t,\rho,\theta)=\frac{e^{i(m_j\theta-Et/\hbar)}}{\sqrt{2\pi}}\left(
           \begin{array}{c}
            R_+(\rho) \\
            R_-(\rho) \\
           \end{array}
         \right),
\end{equation}
where the components $R_+(\rho)$ and $R_-(\rho)$ are real radial functions or radial eigenfunctions (but $R_+(\rho)\neq R_-(\rho)$), $m_j=\pm 1/2,\pm 3/2, \pm 5/2,\ldots$ ($m_j=m_j^\pm=\pm\vert m_j\vert$) is the total magnetic quantum number (i.e., our angular quantum number), and $E$ ($E=E^\pm=\pm\vert E^\pm \vert$) is the relativistic total energy of the fermion (or of the system).

Therefore, substituting \eqref{spinor} in \eqref{dirac7}, we obtain the following second-order differential equation for the DE with EDM and PDM in the $(2+1)$-dimensional Minkowski spacetime (i.e., a relativistic hydrogen atom-like equation)
\begin{equation}\label{dirac8}
\left[\frac{d^2}{d\rho^2}+\frac{1}{\rho}\frac{d}{d\rho}-\frac{m^2_s}{\rho^2}+\frac{\rho_0}{\rho}+\left(\frac{E^2-m^2_0 c^4}{\hbar^2 c^2}\right)\right]R_s(\rho)=0,
\end{equation}
where we define
\begin{equation}
m_s\equiv\left(\xi-\frac{s}{2}\right)>0, \ \ \xi\equiv\sqrt{\left(m_j-\frac{d\lambda_m}{\hbar}\right)^2+\left(\frac{\kappa c}{\hbar}\right)^2}> 0, \ \  \rho_0\equiv\left(\frac{2m_0 c^2\kappa}{\hbar^2}\right)\geq 0,
\end{equation}
being $\xi$ a positive real quantity ($\xi>s/2$) emerged from $\Gamma^2\phi=\xi^2\phi$ and $\Gamma\phi=\pm\xi\phi=s\xi\phi$, where the parameter $s=\pm 1$ (spinorial parameter) describes the components of the spinor, i.e., $s=+1$ is for the upper component and $s=-1$ is for the lower component, respectively.

\section{Relativistic Bound-state solutions: Dirac spinor and the relativistic energy spectrum\label{sec3}}

To solve exactly and analytically Eq. \eqref{dirac8}, let us first introduce a new variable into the system and, preferably, that it has no dimension (i.e., a dimensionless variable). In particular, a good choice for this is given by: $z=2\eta\rho\geq 0$, where $\eta\equiv\sqrt{m^2_0 c^4-E^2}/\hbar c$, being $m^2_0 c^4\geq E^2$ or $\pm m_0 c^2\geq E$ (i.e., a condition for the existence of relativistic bound states \cite{O10,O11,O12,Greiner,Strange}). In that way, through a change of variable, Eq. \eqref{dirac8} becomes
\begin{equation}\label{dirac9}
\left[\frac{d^2}{dz^2}+\frac{1}{z}\frac{d}{dz}-\frac{m^2_s}{z^2}+\frac{z_0}{z}-\frac{1}{4}\right]R_s(z)=0,
\end{equation}
where we define
\begin{equation}
z_0\equiv\frac{\rho_0}{2\eta}.
\end{equation}

Now, we need to analyze the asymptotic behavior (or asymptotic limit) of Eq. \eqref{dirac9} for $z\to 0$ and $z\to\infty$. So, making this, we obtain a regular solution to this equation, such as
\begin{equation}\label{solution}
R_s(z)=C_s z^{m_s}e^{-z/2}\bar{R}_s(z), \ \ (m_s=\vert m_s\vert>0),
\end{equation}
where $C_s>0$ are normalization constants, $\bar{R}_s(z)$ are unknown functions to be determined, and $R_s(z)$ must satisfy the following boundary conditions to be a normalizable solution
\begin{equation}\label{conditions} 
R_s(z\to 0)=0, \ \ R_s(z\to\infty)=0.
\end{equation}

So, substituting \eqref{solution} in \eqref{dirac9}, we obtain a second-order differential equation for $\bar{R}_s(z)$, given as follows
\begin{equation}\label{dirac10}
\left[z\frac{d^{2}}{dz^{2}}+(b_s-z)\frac{d}{dz}-a_s\right]\bar{R}_s(z)=0,
\end{equation}
where
\begin{equation}\label{define}
b_s\equiv 2m_s+1, \ \ a_s\equiv\frac{b_s}{2}-z_0.
\end{equation}

It is not difficult to note that Eq. \eqref{dirac10} is the well-known generalized Laguerre equation, whose solutions are the generalized Laguerre polynomials, written as $\bar{R}_s(z)=L^{2m_s}_n(z)$  \cite{O10,O11,O12,Greiner}. Consequently, the parameter $a_s$ must be equal to a negative integer, i.e., $a_s=-n$ (a quantization condition), where $n=n_\rho=0,1,2,\ldots$ is a quantum number (sometimes called radial quantum number). Therefore, from the condition $a_s=-n$, we obtain the following relativistic energy spectrum for Dirac fermions with EDM and PDM in the presence of a magnetic field generated by magnetic monopoles (i.e., a relativistic hydrogen atom-like spectrum)
\begin{equation}\label{spectrum}
E^\pm_{n,m_j,s}=\pm m_0c^2\sqrt{1-\left[\frac{\kappa c}{\hbar\left(n+\frac{1-s}{2}+\sqrt{\left(m_j-\frac{d\lambda_m}{\hbar}\right)^2+\left(\frac{\kappa c}{\hbar}\right)^2}\right)}\right]^2},
\end{equation}
where the positive signal ($+$) describes the positive-energy states or solutions (i.e., a fermion or particle with a spectrum $E_{particle}=E^+>0$), and the negative signal ($-$) describes the negative-energy states or solutions (i.e., an antifermion or antiparticle with a spectrum $E_{antiparticle}=-E^-=\vert E^-\vert>0$), respectively. That is, as a particle with negative energy (given by $E^-<0$) is actually an antiparticle with positive energy \cite{Greiner,Strange,Martin,griffiths}, we then write the spectrum of the antiparticle in modulus or absolute values. So, we see that the spectrum \eqref{spectrum} is quantized (discretized) in terms of the quantum numbers $n$ and $m_j$, and explicitly depends on the EDM $d$, PDM parameter $\kappa$, magnetic charge density $\lambda_m$, and on the spinorial parameter $s$, respectively. Additionally, a noteworthy observation is that the particle and antiparticle possess equal energies ($E_{particle}=E_{antiparticle}$), i.e., here the spectrum is symmetrical (the system does not distinguish between particles and antiparticles). Consequently, the energy levels exhibit equal spacing on both sides of $E=0$ (this is just a reference point). Therefore, this symmetry in energy levels emphasizes the equilibrium between particle and their respective antiparticle in our system (or in other words, if the pair creation phenomenon were applied here, a gamma photon would give up half of its energy to create a particle and the other half to create the antiparticle). In particular, the quantization of the spectrum is a direct result of the presence of $\kappa$, where such parameter acts as a kind of ``2D Coulomb field or potential'' \cite{O1,Alhaidari1,Alhaidari2,Vakarchuk,O10,O11,O12}. Therefore, here we technically do not have a free fermion (whose spectrum in this case is continuous), but subject to a ``field or potential'' modeled by $\kappa$. Besides, as the condition $\pm m_0 c^2\geq E$ (or $\eta\geq 0)$ must be obeyed for there to be bound states here, it implies that the maximum value for the spectrum $E$ is the rest energy $\pm m_0 c^2$ (i.e., $E_{max}=E_0=\pm m_0 c^2$), which is precisely the rest energy of the particle and antiparticle. In fact, to achieve this it is necessary to take the following limits: $\kappa\to 0$, that is the limit of constant mass, $\lambda_m\to\infty$, that is the limit of very high or infinite density of magnetic charge, and $n=m_j\to\infty$, that is the limit of very large quantum numbers (Bohr's Correspondence Principle) or the limit of continuous. It is also interesting to comment that the presence of $d$ and $\lambda_m$ in the spectrum allows the appearance of a quantum phase of a topological nature, that is, we are talking about the He-McKellar-Wilkens (HMW) topological phase, given by $\varphi_{HMW}=-\frac{4\pi d\lambda_m}{\hbar c}$, and is basically a result of the HMW effect \cite{Dowling}. Consequently, the spectrum \eqref{spectrum} with $-\frac{d\lambda_m}{\hbar}\to\frac{c\varphi_{HMW}}{4\pi}$ takes a very similar form to the spectrum of a Dirac fermion with magnetic dipole moment (MDM) and PDM in the presence of the electric field of the Aharonov-Casher (AC) effect \cite{O12}.

On the other hand, it is interesting to analyze the spectrum according to the values $m_j$, $s$, and $\sigma$, where $\sigma=\pm 1$ (came from $d=\sigma\vert d\vert$) is a parameter (EDM parameter) that describes a positive or negative EDM, i.e., $\sigma=+1$ is for a positive EDM and $\sigma=-1$ is for a negative EDM, respectively. Therefore, in Table \eqref{tab} we have eight possible settings for the spectrum depending on the values of $m_j$, $s$, and $\sigma$. So, according to this Table, we see that some spectra or settings are equal, that is, setting 1 = setting 6; setting 2 = setting 5; setting 3 = setting 8; and setting 4 = setting 7, where the setting 4 (or 7) represents the spectrum with higher or maximum energy (large denominator), and the setting 1 (or 6) represents the spectrum with lower or minimum energy (small denominator), respectively. Therefore, the energies are greater when the particle (or antiparticle) has positive angular momentum and negative EDM ($m_j>0$ and $\sigma=-1$) or negative angular momentum and positive EDM ($m_j<0$ and $\sigma=+1$), and the energies are smaller when the particle (or antiparticle) has positive angular momentum and positive EDM ($m_j>0$ and $\sigma=+1$) or negative angular momentum and negative EDM ($m_j<0$ and $\sigma=-1$), respectively.
\begin{table}[h]
\centering
\begin{small}
\caption{Relativistic spectrum depends on the values of $m_j$, $s$, and $\sigma$.} \label{tab}
\begin{tabular}{ccc}
\hline
Setting & $(m_j, s, \sigma)$ & Relativistic spectrum \\
\hline
1& $(m_j>0,s=+1,\sigma=+1)$ & \ \ $E^\pm=\pm m_0c^2\sqrt{1-\left[\frac{\kappa c}{\hbar\left(n+\sqrt{\left(m_j-\frac{\vert d\vert\lambda_m}{\hbar}\right)^2+\left(\frac{\kappa c}{\hbar}\right)^2}\right)}\right]^2}$\\
2& $(m_j>0,s=+1,\sigma=-1)$ & \ \ $E^\pm=\pm m_0c^2\sqrt{1-\left[\frac{\kappa c}{\hbar\left(n+\sqrt{\left(m_j+\frac{\vert d\vert\lambda_m}{\hbar}\right)^2+\left(\frac{\kappa c}{\hbar}\right)^2}\right)}\right]^2}$\\
3& $(m_j>0,s=-1,\sigma=+1)$ & \ \ $E^\pm=\pm m_0c^2\sqrt{1-\left[\frac{\kappa c}{\hbar\left(n+1+\sqrt{\left(m_j-\frac{\vert d\vert\lambda_m}{\hbar}\right)^2+\left(\frac{\kappa c}{\hbar}\right)^2}\right)}\right]^2}$\\
4& $(m_j>0,s=-1,\sigma=-1)$ & \ \ $E^\pm=\pm m_0c^2\sqrt{1-\left[\frac{\kappa c}{\hbar\left(n+1+\sqrt{\left(m_j+\frac{\vert d\vert\lambda_m}{\hbar}\right)^2+\left(\frac{\kappa c}{\hbar}\right)^2}\right)}\right]^2}$\\
5& $(m_j<0,s=+1,\sigma=+1)$ & \ \ $E^\pm=\pm m_0c^2\sqrt{1-\left[\frac{\kappa c}{\hbar\left(n+\sqrt{\left(\vert m_j\vert+\frac{\vert d\vert\lambda_m}{\hbar}\right)^2+\left(\frac{\kappa c}{\hbar}\right)^2}\right)}\right]^2}$\\
6& $(m_j<0,s=+1,\sigma=-1)$ & \ \ $E^\pm=\pm m_0c^2\sqrt{1-\left[\frac{\kappa c}{\hbar\left(n+\sqrt{\left(\vert m_j\vert-\frac{\vert d\vert\lambda_m}{\hbar}\right)^2+\left(\frac{\kappa c}{\hbar}\right)^2}\right)}\right]^2}$\\
7& $(m_j<0,s=-1,\sigma=+1)$ & \ \ $E^\pm=\pm m_0c^2\sqrt{1-\left[\frac{\kappa c}{\hbar\left(n+1+\sqrt{\left(\vert m_j\vert+\frac{\vert d\vert\lambda_m}{\hbar}\right)^2+\left(\frac{\kappa c}{\hbar}\right)^2}\right)}\right]^2}$\\
8& $(m_j<0,s=-1,\sigma=-1)$ & \ \ $E^\pm=\pm m_0c^2\sqrt{1-\left[\frac{\kappa c}{\hbar\left(n+1+\sqrt{\left(\vert m_j\vert-\frac{\vert d\vert\lambda_m}{\hbar}\right)^2+\left(\frac{\kappa c}{\hbar}\right)^2}\right)}\right]^2}$\\
\hline
\end{tabular}
\end{small}
\end{table}

\begin{figure}[!h]
\centering
\includegraphics[width=10.0cm]{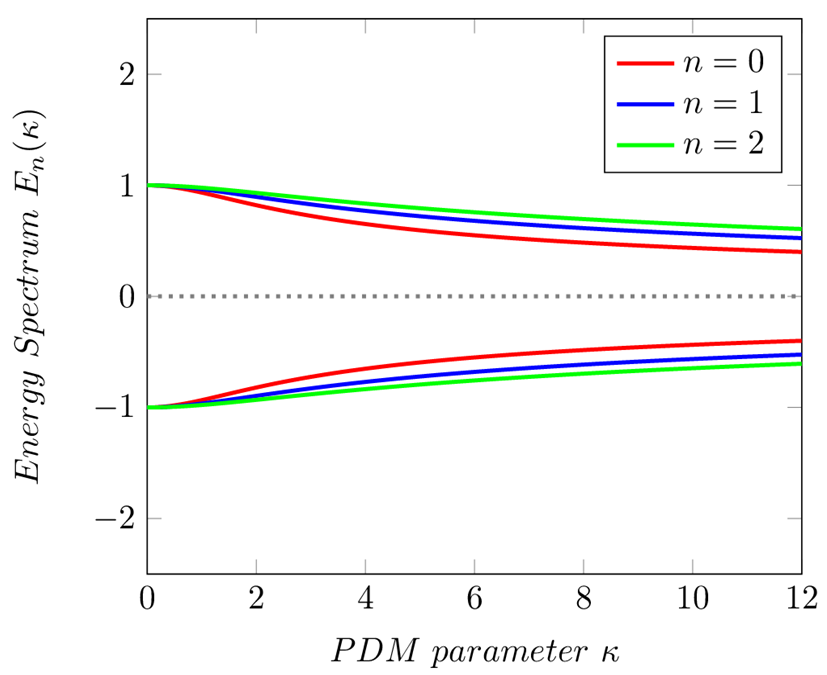}
\caption{Behavior of $E_n(\kappa)$ versus $\kappa$ for three different values of $n$.}
\label{fig1}
\end{figure}

\begin{figure}[!h]
\centering
\includegraphics[width=10.0cm]{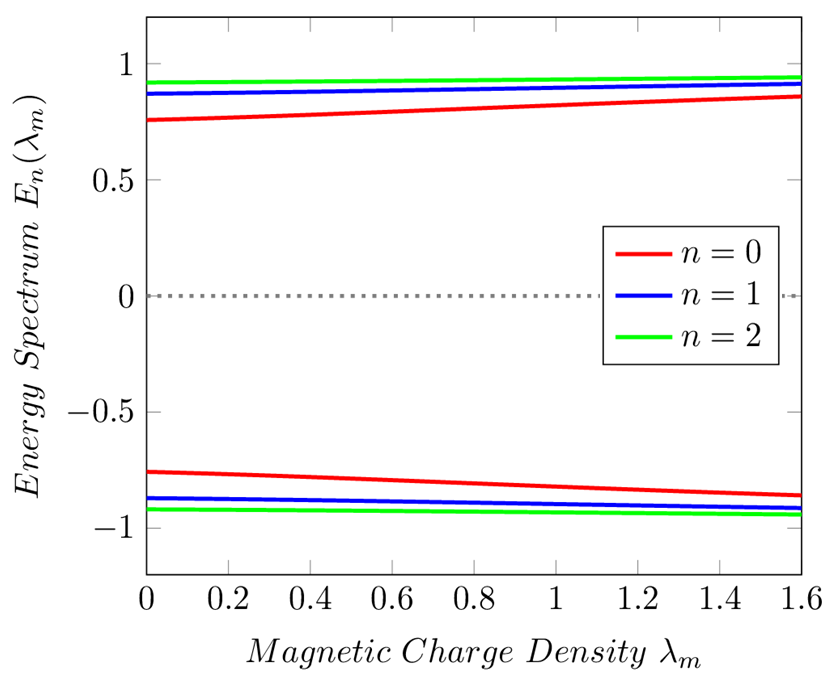}
\caption{Behavior of $E_n(\lambda_m)$ versus $\lambda_m$ for three different values of $n$.}
\label{fig2}
\end{figure}

Now, let us graphically analyze the behavior of the spectrum (or energy levels) as a function of the PDM parameter $\kappa$ and of the magnetic charge density $\lambda_m$ for three different values of $n$ (with $m_j=1/2$, i.e., we consider the spectrum of setting 4, which is the spectrum with the highest values). For simplicity, here we adopt the natural unit system ($\hbar=m_0=1$), and a unitary EDM ($\vert d\vert=1$). In this way, in Fig. \ref{fig1} we have the behavior of the spectrum $E_n(\kappa)$ as a function of $\kappa$ for the ground state ($n = 0$) and the first two excited states ($n=1,2$), where we consider $\lambda_m=1$. So, according to this Figure, we see that the energies increase with the increase of $n$ (as it should be, otherwise, something would be wrong), that is, the energy difference between two consecutive levels is positive ($\Delta E_n=E_{n+1}-E_n>0$). Furthermore, the energies decrease with the increase of $\kappa$, that is, the goal of $\kappa$ is to reduce the energies of the particle (or antiparticle). For example, in the limit $\kappa\to\infty$, we have $E_n\to 0$. Therefore, the more massive the particle (or antiparticle), the less energy it has (a ``paradox''). Already in Fig. \ref{fig2}, we have the behavior of the spectrum $E_n(\lambda_m)$ as a function of $\lambda_m$ for the ground state ($n = 0$) and the first two excited states ($n=1,2$), where we consider $\kappa=2$. So, according to this Figure, we see that the energies also increase with the increase of $n$ (as it should be), where the energy difference between two consecutive levels is also positive ($\Delta E_n>0$). Furthermore, the energies increase with the increase of $\lambda_m$, that is, the goal of $\lambda_m$ is to increase the energies of the particle (or antiparticle). Therefore, the more magnetic monopoles the system has, the more energy the particle (or antiparticle) has.

From here on, let us concentrate our attention on the form of the two-component Dirac spinor for the relativistic bound states of the system. So, using the variable $z=2\eta\rho$ in \eqref{solution}, we obtain
\begin{equation}\label{solution2}
R_s(\rho)=C_s (2\eta)^{m_s} \rho^{m_s}e^{-\eta\rho}L_n^{2m_s}(2\eta\rho), \ \ (m_s=\vert m_s\vert>0).
\end{equation}

In that way, the spinor \eqref{spinor} becomes
\begin{equation}\label{spinor2}
\phi(t,\rho,\theta)=e^{i(m_j\theta-Et/\hbar)}\left(
           \begin{array}{c}
            \Bar{C}_+\rho^{m_+}e^{-\eta\rho}L_n^{2m_+}(2\eta\rho) \\
            \Bar{C}_-\rho^{m_-}e^{-\eta\rho}L_n^{2m_-}(2\eta\rho) \\
           \end{array}
         \right),
\end{equation}
where $\Bar{C}_s\equiv\frac{C_s (2\eta)^{m_s}}{\sqrt{2\pi}}$ is a new normalization constant.

However, as $\phi$ is given by $\phi=U^\dagger \Phi$, implies that $\Phi=(U^\dagger)^{-1}\phi=U\phi$. Therefore, knowing that $\Psi_D=O_D\Phi$, where $O_D$ is actually the operator $A^+$, given in \eqref{dirac5}, we can write $\Psi_D$ as
\begin{equation}\label{spinor3}
\Psi_D(t,\rho,\theta)=\left[\gamma^0\left(\frac{i\hbar}{c}\frac{\partial}{\partial t}\right)+i\gamma^\rho\left(\hbar\frac{\partial}{\partial\rho}+d\gamma^0 B(\rho)\right)+\gamma^\theta\left(\frac{i\hbar}{\rho}\frac{\partial}{\partial\theta}\right)+ m(\rho)c\right]U\phi,
\end{equation}
or better
\begin{equation}\label{spinor4}
\Psi_D(t,\rho,\theta)=\left[\gamma^0\left(\frac{i\hbar}{c}\frac{\partial}{\partial t}\right)+i\gamma^\rho\left(\hbar\frac{\partial}{\partial\rho}+\frac{d\lambda_m}{\rho}\gamma^0\right)+\gamma^\theta\left(\frac{i\hbar}{\rho}\frac{\partial}{\partial\theta}\right)+\left(m_0 c-\frac{\kappa c}{\rho}\right)\right]U\phi,
\end{equation}
where we use \eqref{magneticfield} and \eqref{variablemass}.

On the other hand, to move forward, we must convert $\gamma^\rho$ and $\gamma^\theta$ in $\gamma^1$ and $\gamma^2$, in which it is made via similarity transformation. So, by doing this in \eqref{spinor4}, we have
\begin{equation}\label{spinor5}
\Psi_D(t,\rho,\theta)=U\left\{U^{\dagger}\left[\gamma^0\left(\frac{i\hbar}{c}\frac{\partial}{\partial t}\right)+i\gamma^\rho\left(\hbar\frac{\partial}{\partial\rho}+\frac{d\lambda_m}{\rho}\gamma^0\right)+\gamma^\theta\left(\frac{i\hbar}{\rho}\frac{\partial}{\partial\theta}\right)+\left(m_0 c-\frac{\kappa c}{\rho}\right)\right]U\right\} \phi,
\end{equation}
where implies
\begin{equation}\label{spinor6}
\Psi_D(t,\rho,\theta)=UB^+\phi=U\left[\sigma_3\left(\frac{i\hbar}{c}\frac{\partial}{\partial t}\right)+\sigma_2\left(\hbar\frac{\partial}{\partial\rho}+\frac{\hbar}{2\rho}\right)+i\sigma_1\left(\frac{i\hbar}{\rho}\frac{\partial}{\partial\theta}+\frac{d\lambda_m}{\rho}\right)+\left(m_0 c-\frac{\kappa c}{\rho}\right)\right]\phi,
\end{equation}
that is, we obtain the operator $B^+$, given by \eqref{operators4}. Explicitly, this operator is written as follows
\begin{equation}
B^+=\left(
    \begin{array}{cc}
      \left(\frac{i\hbar}{c}\frac{\partial}{\partial t}\right)+\left(m_0 c-\frac{\kappa c}{\rho}\right)\ &  -i\left(\hbar\frac{\partial}{\partial\rho}+\frac{\hbar}{2\rho}\right)+i\left(\frac{i\hbar}{\rho}\frac{\partial}{\partial\theta}+\frac{d\lambda_m}{\rho}\right) \\
      i\left(\hbar\frac{\partial}{\partial\rho}+\frac{\hbar}{2\rho}\right)+i\left(\frac{i\hbar}{\rho}\frac{\partial}{\partial\theta}+\frac{d\lambda_m}{\rho}\right)\ & -\left(\frac{i\hbar}{c}\frac{\partial}{\partial t}\right)+\left(m_0 c-\frac{\kappa c}{\rho}\right) \\
    \end{array}
  \right).
\end{equation}

Therefore, replacing \eqref{spinor2} in \eqref{spinor6}, we obtain the following two-component Dirac spinor for the relativistic bound states of the system
\begin{equation}\label{spinor7}
\Psi_D(t,\rho,\theta)=e^{-iEt/\hbar}\left(
           \begin{array}{c}
            e^{i(m_j-1/2)\theta}[F_+-iG_-] \\
            e^{i(m_j+1/2)\theta}[F_- +iG_+] \\
           \end{array}
         \right),
\end{equation}
where we define
\begin{equation}\label{define1}
F_s(\rho)\equiv\Bar{C}_s\left(s\frac{E}{c}+m_0 c-\frac{\kappa c}{\rho}\right)\rho^{m_s}e^{-\eta\rho}L_{n}^{2m_s}(2\eta\rho), \ \ (s=\pm 1),
\end{equation}
and
\begin{equation}\label{define2}
G_s(\rho)\equiv\hbar\Bar{C}_s\rho^{m_s}e^{-\eta\rho}\left[L_{n}^{2m_s}(2\eta\rho)\left(\frac{m_s-sm_j+\frac{1}{2}+s\frac{d\lambda_m}{\hbar}}{\rho} \right)-2\eta L_{n-1}^{2m_s+1}(2\eta\rho)\right],
\end{equation}
being $U=e^{-i\theta\sigma_3/2}$=diag$(e^{-i\theta/2},e^{+i\theta/2})$ and we use the fact that $\frac{d}{d\rho}L_n^{2m_s}(2\eta\rho)=-2\eta L_{n-1}^{2m_s+1}(2\eta\rho)$.

\section{Nonrelativistic Bound-state solutions: wave function and the nonrelativistic energy spectrum\label{sec4}}

To analyze the nonrelativistic limit (low-energy regime) of our results, it is necessary to consider that most of the total energy of the system is concentrated in the rest energy of the particle \cite{Greiner,Strange,Martin,O10,O11,O12}. In this way, it needs to consider that $E=m_0 c^2+\varepsilon$ (with $s=+1$), where $m_0 c^2\gg \varepsilon$, being $\varepsilon$ the nonrelativistic spectrum, as well as $\kappa\ll 1$ or $(\kappa c/\hbar)\ll 1$ (small effective mass). Therefore, applying these conditions in Eq. \eqref{dirac8} (with $R_+(\rho)\to \psi_S(\rho)$), we obtain the following 2D SE for a spinless particle with PDM and EDM (i.e., a nonrelativistic hydrogen atom-like equation)
\begin{equation}\label{non1}
\left[-\frac{\hbar^2}{2m_0}\left(\frac{d^2}{d\rho^2}+\frac{1}{\rho}\frac{d}{d\rho}-\frac{l^2}{\rho^2}\right)-\frac{c^2\kappa}{\rho}\right]\psi_S(\rho)=\varepsilon \psi_S(\rho),
\end{equation}
or in time-dependent form, such as
\begin{equation}\label{non2}
\left[-\frac{\hbar^2}{2m_0}\left(\frac{\partial^2}{\partial\rho^2}+\frac{1}{\rho}\frac{\partial}{\partial\rho}-\frac{\left(L_z-\frac{d\lambda_m}{\hbar}\right)^2}{\rho^2}\right)-\frac{c^2\kappa}{\rho}\right]\Psi_S(t,\rho,\theta)=i\hbar\frac{\partial}{\partial t} \Psi_S(t,\rho,\theta),
\end{equation}
where $l\equiv m-\frac{d\lambda_m}{\hbar}$, $m\equiv m_j-1/2=0,\pm 1,\pm 2,\pm 3,\ldots$ are the eigenvalues of $L_z=-i\hbar\frac{\partial}{\partial\theta}$ (i.e., $m$ is the orbital magnetic quantum number or simply magnetic quantum number), and $\Psi_S(t,\rho,\theta)=\frac{e^{i(m\theta-\varepsilon t/\hbar)}}{\sqrt{2\pi}}\psi_S(\rho)$ is the Schrödinger wave function. As we can see, this wave function is similar to the upper component of the spinor \eqref{spinor}, where $m$ (an integer) plays the role of $m_j$ (a half-integer) and $\varepsilon$ plays the role of $E$, respectively. Besides, to find the form of the function $\psi_S(\rho)$, we can solve Eq. \eqref{non1} just as we solved Eq. \eqref{dirac8}. However, an easier and more direct way is to make the nonrelativistic limit of \eqref{solution2}. Therefore, doing $R_+ (\rho)\to\psi_S (\rho)$, $\eta\to\bar{\eta}=\sqrt{-2m_0\varepsilon}/\hbar>0$ (with $\varepsilon<0$), $C_+\to C_S$, $m_+\to\vert l\vert$ in \eqref{solution2}, $\psi_S (\rho)$ takes the form
\begin{equation}
\psi_S (\rho)=C_S(2\bar{\eta})^{\vert l\vert} \rho^{\vert l\vert}e^{-\bar{\eta}\rho}L_n^{2\vert l\vert}(2\bar{\eta}\rho),
\end{equation}
and, consequently, the wave function is written as
\begin{equation}
\Psi_S (t,\rho,\theta)=\bar{C}_S e^{i(m\theta-\varepsilon t/\hbar)}\rho^{\vert l\vert}e^{-\bar{\eta}\rho}L_n^{2\vert l\vert}(2\bar{\eta}\rho),
\end{equation}
where we define $\bar{C}_S\equiv\frac{C_S(2\bar{\eta})^{\vert l\vert} }{\sqrt{2\pi}}$.

Now, let us focus our attention on the form of the nonrelativistic spectrum of the system. So, defining $x\equiv(\kappa c/\hbar)^2$ in \eqref{spectrum} \cite{O10,O11,O12}, we can obtain from $E=m_0 c^2+\varepsilon$ the following nonrelativistic spectrum (with $m_j\to m+1/2$)
\begin{equation}\label{varepsilon}
\varepsilon_{n,m}=m_0c^2\left[f(x)-1\right],
\end{equation}
where
\begin{equation}
f(x)\equiv \sqrt{1-\frac{x}{\left(n+\sqrt{\left(m+\frac{1}{2}-\frac{d\lambda_m}{\hbar}\right)^2+x}\right)^2}}.
\end{equation}

So, expanding $f(x)$ into a first-order Taylor series, we obtain
\begin{equation}\label{Taylor}
f(x)=f(x)+x\frac{df(x)}{dx}=1-\frac{1}{2}\frac{x}{\left(n+m+\frac{1}{2}-\frac{d\lambda_m}{\hbar}\right)^2}=1-\frac{1}{2}\frac{(\kappa c/\hbar)^2}{\left(n+m+\frac{1}{2}-\frac{d\lambda_m}{\hbar}\right)^2}.
\end{equation}

\begin{figure}[!h]
\centering
\includegraphics[width=10.0cm]{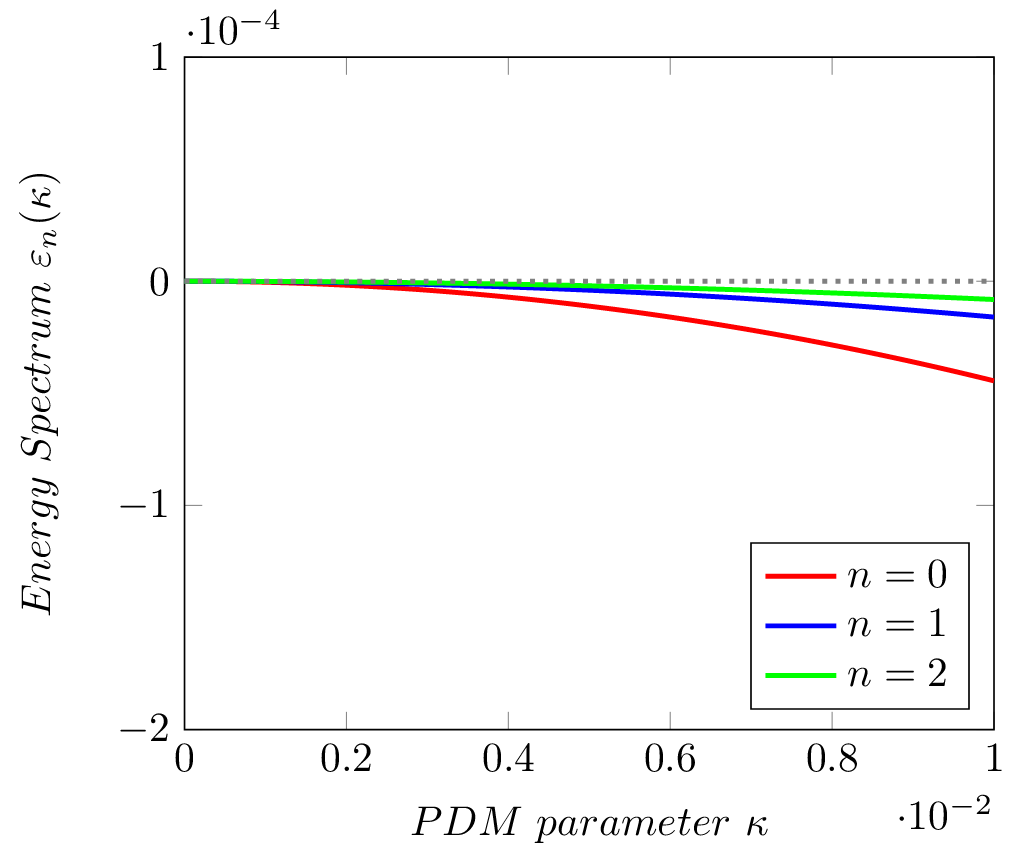}
\caption{Behavior of $\varepsilon_n(\kappa)$ versus $\kappa$ for three different values of $n$.}
\label{fig3}
\end{figure}

\begin{figure}[!h]
\centering
\includegraphics[width=10.0cm]{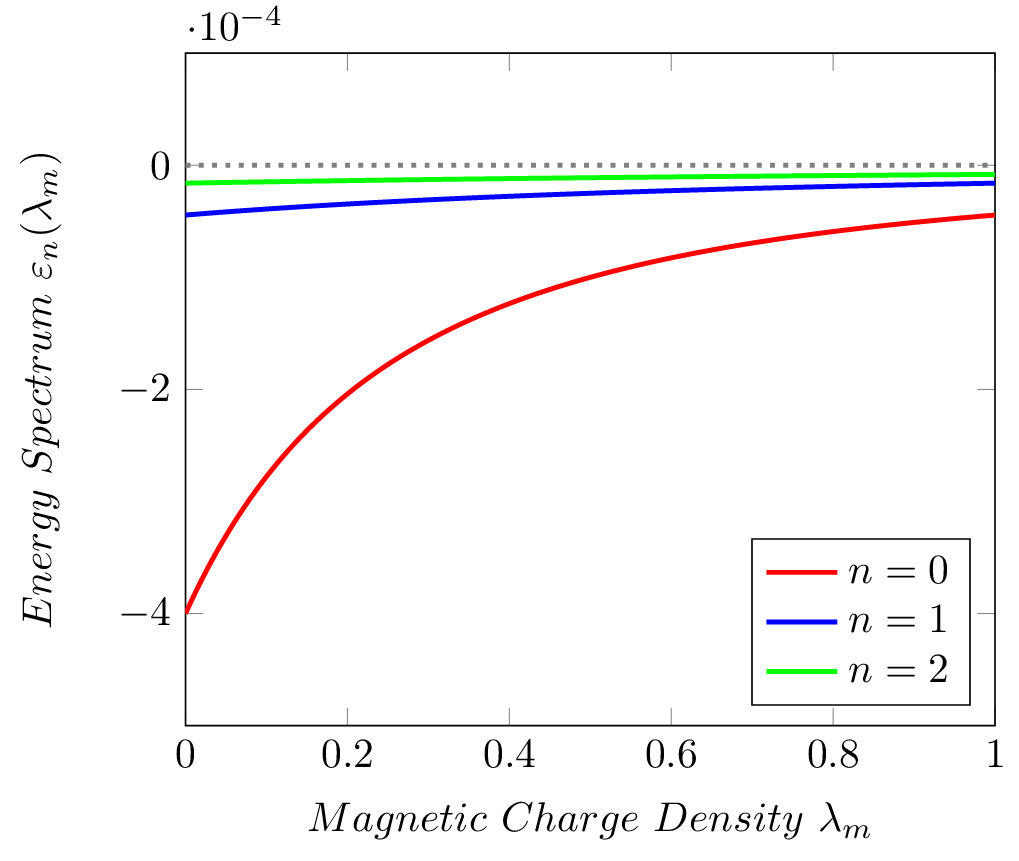}
\caption{Behavior of $\varepsilon_n(\lambda_m)$ versus $\lambda_m$ for three different values of $n$.}
\label{fig4}
\end{figure}

Therefore, from \eqref{Taylor} and \eqref{varepsilon}, we obtain the following nonrelativistic spectrum for a spinless particle with PDM and EDM in the presence of a magnetic field generated by magnetic monopole (i.e., a nonrelativistic hydrogen atom-like spectrum)
\begin{equation}\label{spectrum2}
\varepsilon_{n,m}=-\frac{1}{2}\frac{m_0 c^4 \kappa^2}{\hbar^2\left(n+m+\frac{1}{2}-\frac{d\lambda_m}{\hbar}\right)^2}, \ \ (n=0,1,2,3\ldots; \ m=0,\pm 1,\pm 2,\pm 3,\ldots),
\end{equation}
where $\left(n+m+\frac{1}{2}-\frac{d\lambda_m}{\hbar}\right)\neq 0$ or $\frac{d\lambda_n}{\hbar}\neq\frac{1}{2}$ (otherwise the spectrum diverges to $n=m=0$). In particular, the negative sign in the spectrum does not have the same meaning in the relativistic case (antiparticle states). Here, it means that the energies must increase as quantum numbers increase, such as occurs in the relativistic case for the particle (and also in the nonrelativistic hydrogen atom \cite{Greiner}). So, still analogous to the relativistic case (for the particle), we see that the values of the spectrum \eqref{spectrum2} decreases with increasing $\kappa$, where $\kappa$ acts as a kind of ``2D Coulomb field or potential'', {\color{red}and} increase with the increase of $\lambda_m$ (for $d<0$), respectively. Besides, with respect to the constant $c^4$ (speed of light raised to the fourth power) that appears explicitly in \eqref{spectrum2}, such constant is canceled due to the presence of $\kappa$ (since $\kappa \sim 1/c^2$). In Fig. \ref{fig3}, we have the behavior of the spectrum $\varepsilon_n (\kappa)$ as a function of $\kappa$ for the ground state ($n=0$) and the first two excited states ($n=1,2$), where we consider $\hbar=m_0=\lambda_m=1$, $m=0$ and $d=-1$ (such as in the relativistic case for the particle). Already in Fig. \ref{fig4}, we have the behavior of the spectrum $\varepsilon_n (\lambda_m)$ as a function of $\lambda_m$ for $n=0,1,2$, where we consider $\hbar=m_0=1$, $m=0$, $d=-1$ and $\kappa=0.01=1/100$ (since $\kappa\ll 1$).

\section{Conclusions}\label{sec5}

In this paper, we determine to determine the relativistic and nonrelativistic bound-state solutions for Dirac fermions with EDM and PDM in the presence of a magnetic field generated by magnetic monopoles. To achieve this goal, we work with the $(2+1)$-dimensional DE with nonminimal coupling in polar coordinates $(t,\rho,\theta)$, where the coupling constant is the EDM of the fermion. To solve exactly and analytically the DE, we first convert it to a quadratic DE (i.e., a second-order differential equation). In that way, we obtain the bound-state solutions of the system, given by the two-component Dirac spinor and by the relativistic energy spectrum. With respect to the magnetic field, we consider the radial magnetic field generated by a straight line (or linear distribution) of magnetic charges, given by: $\Vec{B}=\frac{\lambda_m}{\rho}\Vec{e}_\rho$, being $\lambda_m$ the magnetic charge density. Now, with respect to the PDM of our problem, we consider a PDM given by: $m(\rho)=m_0-\frac{\kappa}{\rho}$, being $m_0$ the rest mass and $\kappa$ a positive real parameter (PDM parameter).

So, after solving the second-order differential equation via a change of variable and asymptotic behavior, we obtain the bound-state solutions of the system, where we note that the Dirac spinor is written in terms of the generalized Laguerre polynomials, and the spectrum (for a fermion/particle and an antifermion/antiparticle) is quantized (discretized) in terms of the radial and total magnetic quantum numbers $n=0,1,2,\ldots$ and $m_j=\pm 1/2,\pm 3/2,\pm 5/2,\ldots$, and explicitly depends on the EDM $d$, PDM parameter $\kappa$, magnetic charge density $\lambda_m$, and on the spinorial parameter $s$, respectively. Additionally, a noteworthy observation is that the particle and antiparticle possess equal energies, i.e., here the spectrum is symmetrical. Consequently, the
energy levels exhibit equal spacing on both sides of $E=0$. Therefore, this symmetry in energy levels emphasizes the equilibrium between particle and their respective antiparticle in our system (in other words, if the pair creation phenomenon were applied here, a gamma photon would give up half of its energy to create a particle and the other half to create the antiparticle). 

In particular, the quantization of the spectrum is a direct result of the presence of $\kappa$, where such parameter acts as a kind of ``2D Coulomb field or potential''. Therefore, here we technically do not have a free fermion (whose spectrum in this case is continuous), but subject to a ``field or potential'' modeled by $\kappa$. Besides, as the condition $\pm m_0 c^2\geq E$ (condition for bound states), it implies that the maximum value for the spectrum $E$ is the rest energy $\pm m_0 c^2$ (i.e., $E_{max}=E_0=\pm m_0 c^2$), which is precisely the rest energy of the particle and antiparticle. In fact, to achieve this it is necessary to take the following limits: $\kappa\to 0$, that is the limit of constant mass, $\lambda_m\to\infty$, that is the limit of very high or infinite density of magnetic charge, and $n=m_j\to\infty$, that is the limit of very large quantum numbers (Bohr's Correspondence Principle) or the limit of continuous. It is also interesting to comment that the presence of $d$ and $\lambda_m$ in the spectrum allows the appearance of a quantum phase of a topological nature: the He-McKellar-Wilkens (HMW) topological phase (of the HMW effect). Consequently, our spectrum takes a very similar form to the spectrum of a Dirac fermion with magnetic dipole moment (MDM) and PDM in the Aharonov-Casher (AC) effect.

Furthermore, we also analyze the spectrum according to the values $m_j$, $s$, and $\sigma$, where $\sigma=\pm 1$ is a parameter that describes a positive or negative EDM. So, from this, we obtain a Table with eight possible settings for the spectrum. For example, according to this Table, we see that the energies are greater when the particle (or antiparticle) has positive angular momentum and negative EDM ($m_j>0$ and $\sigma=-1$) or negative angular momentum and positive EDM ($m_j<0$ and $\sigma=+1$), and the energies are smaller when the particle (or antiparticle) has positive angular momentum and positive EDM ($m_j>0$ and $\sigma=+1$) or negative angular momentum and negative EDM ($m_j<0$ and $\sigma=-1$), respectively. By way of illustration, we also graphically analyzed the behavior of the spectrum $E_n$ as a function of $\kappa$ and $\lambda_m$ for three different values of $n$ (with $m_j=1/2$), i.e., for the ground state ($n=0$), first excited state ($n=1$), and second excited state ($n=2$). In that way, in graph $E_n(\kappa)$ versus $\kappa$, we see that the energies increase with the increase of $n$ (as it should be), and the energies decrease with the increase of $\kappa$, that is, the goal of $\kappa$ is to reduce the energies of the particle (or antiparticle). For example, in the limit $\kappa\to\infty$, we have $E_n\to 0$. Therefore, the more massive the particle (or antiparticle), the less energy it has. Already in graph $E_n(\lambda_m)$ versus $\lambda_m$, we see that the energies also increase with the increase of $n$ (as it should be), and the energies increase with the increase of $\lambda_m$, that is, the goal of $\lambda_m$ is to increase the energies of the particle (or antiparticle). Therefore, the more magnetic monopoles the system has, the more energy the particle (or antiparticle) has.

Finally, we also analyze the nonrelativistic limit (low-energy regime) of our results. For this, we had to consider that most of the total energy of the system is concentrated in the rest energy of the particle (i.e., $E=m_0 c^2+\varepsilon$, where $m_0 c^2\gg \varepsilon$)  as well as a small effective mass or PDM ($\kappa\ll 1$). With this, we obtain the 2D Schrödinger equation (SE) for a spinless particle with PDM and EDM. To obtain the solution of this equation, that is, the wave function, we start directly from the nonrelativistic limit of the upper component of the Dirac spinor. Now, with respect to the nonrelativistic aspect of this equation, we obtain via a Taylor series expansion of the relativistic spectrum for the particle. In particular, this spectrum is negative, that is, energies increase with increasing of the quantum numbers $n$ and $m$ as well as of the density of magnetic charge $\lambda_m$, and decreases with increasing of the PDM parameter $\kappa$.

\section*{Acknowledgments}

\hspace{0.5cm}The author would like to thank the Conselho Nacional de Desenvolvimento Cient\'{\i}fico e Tecnol\'{o}gico (CNPq) for financial support.

\section*{Data availability statement}

\hspace{0.5cm} This manuscript has no associated data or the data will not be deposited. [Author’s comment: There is no data because this paper is a theoretical study based on calculations of the relativistic and nonrelativistic bound-state solutions for Dirac fermions with electric dipole moment and position-dependent mass in the presence of a magnetic field generated by magnetic monopoles.]

\end{document}